\documentclass[a4paper,fleqn,usenatbib]{mnras}

\usepackage{mathptmx}

\usepackage[T1]{fontenc}
\usepackage{ae,aecompl}

\usepackage{graphicx}	
\usepackage{amsmath}	
\usepackage{amssymb}	

\title[A glitch in the millisecond pulsar J0613$-$0200]{A glitch in the millisecond pulsar J0613$-$0200}

\author[J.\,W.\,McKee et al.]{J.\,W.\,McKee,$^{1}$\thanks{E-mail: james.mckee@manchester.ac.uk}
G.\,H.\,Janssen,$^{2}$
B.\,W.\,Stappers,$^{1}$
A.\,G.\,Lyne,$^{1}$
R.\,N.\,Caballero,$^{3}$
\newauthor
L.\,Lentati,$^{4}$
G.\,Desvignes,$^{3}$
A.\,Jessner,$^{3}$
C.\,A.\,Jordan,$^{1}$
R.\,Karuppusamy,$^{3}$
M.\,Kramer,$^{1,3}$
\newauthor
I.\,Cognard,$^{5,6}$
D.\,J.\,Champion,$^{3}$
E.\,Graikou,$^{3}$
P.\,Lazarus,$^{3}$
S.\,Os\l{}owski,$^{7,3}$
D.\,Perrodin,$^{8}$
\newauthor
G.\,Shaifullah,$^{3,7}$
C.\,Tiburzi,$^{3,7}$
and J.\,P.\,W.\,Verbiest$^{7,3}$
\\
$^{1}$Jodrell Bank Centre for Astrophysics, School of Physics and Astronomy, The University of Manchester, Manchester M13 9PL, UK\\
$^{2}$ASTRON, the Netherlands Institute for Radio Astronomy, Postbus 2, 7990 AA, Dwingeloo, The Netherlands\\
$^{3}$Max Planck Institut f{\"u}r Radioastronomie, Auf dem H{\"u}gel 69, 53121, Bonn, Germany\\
$^{4}$Institute of Astronomy / Battcock Centre for Astrophysics, University of Cambridge, Madingley Road, Cambridge CB3 0HA, United Kingdom\\
$^{5}$Laboratoire de Physique et Chimie de l'Environnement et de l'Espace LPC2E CNRS-Universit{\'e} d'Orl{\'e}ans, F-45071 Orl{\'e}ans, France\\
$^{6}$Station de radioastronomie de Nan\c{c}ay, Observatoire de Paris, CNRS/INSU F-18330 Nan\c{c}ay, France\\
$^{7}$Fakult{\"a}t f{\"u}r Physik, Universit{\"a}t Bielefeld, Postfach 100131, 33501 Bielefeld, Germany\\
$^{8}$INAF - Osservatorio Astronomico di Cagliari, via della Scienza 5, I-09047 Selargius (CA), Italy\\
}

\date{Accepted XXX. Received YYY; in original form ZZZ}

\pubyear{2016}

\begin{document}
\label{firstpage}
\pagerange{\pageref{firstpage}--\pageref{lastpage}}
\maketitle

\begin{abstract}
We present evidence for a small glitch in the spin evolution of the
millisecond pulsar J0613$-$0200, using the EPTA Data Release 1.0,
combined with Jodrell Bank analogue filterbank TOAs recorded with the Lovell
telescope and Effelsberg Pulsar Observing System TOAs. A spin frequency step of 0.82(3)\,nHz and frequency
derivative step of ${-1.6(39) \times 10^{-19}\,\text{Hz} \ \text{s}^{-1}}$ are measured
at the epoch of MJD\,$50888(30)$. After PSR\,B1821$-$24A, this is only the second glitch ever
observed in a millisecond pulsar, with a fractional size in frequency of ${\Delta
\nu/\nu=2.5(1) \times 10^{-12}}$, which is several times smaller than the previous smallest glitch. PSR\,J0613$-$0200 is used in gravitational wave searches with
pulsar timing arrays, and is to date only the second such pulsar to
have experienced a glitch in a combined 886 pulsar-years of observations. We find that accurately modelling the glitch does not impact the timing precision for pulsar timing array applications. We
estimate that for the current set of millisecond pulsars included
in the International Pulsar Timing Array, there is a probability of
$\sim 50$\% that another glitch will be observed in a timing array
pulsar within 10 years.
\end{abstract}




\section{Introduction}
Pulsars spin with remarkable stability, allowing pulse times of
arrival (TOAs) to be accurately predicted with precisions, in the best
cases, as high as fractions of microseconds over timescales of
decades. Millisecond pulsars (MSPs) in particular have such
highly stable rotation that
they are used as extremely precise clocks in timing experiments, and
the most stable are used as probes of space-time in pulsar timing
array (PTA) experiments. The ultimate goal is a direct
gravitational wave (GW) detection in the nano-Hertz regime (recent stochastic background limits are given in \citealp{ltm+15}, 
\citealp{abb+15}, \citealp{srl+15}).
Since the influence of GWs on pulse TOAs is extremely small, the
accuracy of the timing model describing the spin evolution of a pulsar needs to be very
high in order to make a GW detection. This also requires the precise measurement and removal of other
influences on the TOAs, such as those caused by changes in the interstellar
medium (ISM) or irregularities in the pulsar spin evolution.

PSR\,J0613$-$0200 was discovered by \cite{lnl+95} and is a MSP which is
included in all currently-ongoing PTA experiments: the European Pulsar
Timing Array (EPTA; \citealp{dcl+16}), the North American Nanohertz Observatory for
Gravitational Waves (NANOGrav; \citealp{abb+15}), the Parkes Pulsar Timing Array (PPTA; \citealp{rhc+16}),
and the International Pulsar Timing Array (IPTA; \citealp{vlh+16}). It has been timed to
a precision of $1.2\,\mu$s over a time span of 13.7 years using the combined IPTA data set (\citealp{vlh+16}).

Although the spin evolution of pulsars is generally very stable and
predictable, a small fraction of pulsars exhibit sudden changes in spin frequency
and/or frequency derivative, known as timing glitches. Timing glitches
are usually associated with non-recycled and low-characteristic-age pulsars, notably the Crab pulsar (PSR\,B0531+21) and the Vela
pulsar (PSR\,B0833$-$45), which have been observed to glitch 25 and 19 times
respectively in 45 years of observations
(\citealp{els+11}). Conversely, glitches in MSPs are exceedingly rare,
with only one small glitch ever observed in 
the MSP B1821$-$24A \citep{cb04}, which is near the core of the globular cluster M28, and which displays significant timing noise (Figure \ref{fig:ppdot}).

The mechanism which causes timing glitches is not fully
understood, but is assumed to be linked to a sudden transfer of
angular momentum from superfluid neutrons to the solid crust. The
superfluid is thought to rotate independently from the rest of the
neutron star and contains vortices. An ensemble of vortices becomes unpinned and a
coupling to the solid component of the neutron star crust occurs, abruptly transferring
angular momentum to it (a review of glitch models can be found in \citealp{hm+15}). The
transfer of angular momentum generally increases the
rotational frequency of the pulsar, which is occasionally observed to relax back to the pre-glitch value, although
\cite{akn+13} have reported evidence for `anti-glitches',
a sudden \textit{decrease} in the spin frequency in X-ray
observations of the magnetar 1E\,2259+586. The change in spin frequency and
slowdown rate caused by a glitch is reflected in the deviation of the observed TOAs
from the arrival times predicted by a pre-glitch timing model.

The glitch observed in PSR\,B1821$-$24A was notable as it was the first
glitch to be observed in a MSP, and had a size ${\Delta \nu/\nu=8(1)
 \times 10^{-12}}$, two orders of magnitude smaller than the next
smallest glitch (at the time). 
The rarity of glitches in MSPs and the small size of the PSR\,B1821$-$24A glitch has led to speculation
that MSPs have different structures to the rest of the population, or a different physical process could be responsible. Some proposed explanations are that
PSR\,B1821$-$24A is a strange star, which experienced a crust-cracking
event that would alter the angular momentum and cause the same
effect on timing residuals as a small glitch \citep{msb+06}, or
that the small size of the glitch is evidence of influence on the pulsar-term by a GW burst with memory,
which could be indistinguishable from a post-glitch frequency step
\citep{cj12}.

Timing noise is a phenomenon where the observed arrival times of
pulses deviate systematically from the timing solution through a process
similar to a random walk in the spin parameters (e.g. \citealp{sc10}). This manifests as structure in the timing residuals. 
Timing noise is thought to arise through unmodelled small-scale instabilities in the rotation of the pulsar. It has been shown that in slow pulsars, timing noise can be modelled as: a series of microglitches \citep{js06}, frequency derivative variations caused by magnetospheric switching \citep{lhk+10}, or as post-glitch recovery stages \citep{hlk10}.

The structure for this paper is as follows: we describe our
observations and data in section 2, present our findings in section 3,
discuss the implications of our results in section 4, and make closing
conclusions in section 5.

\section{Observations}
The data set is comprised of TOAs from a variety of pulsar backends used with
the Lovell Telescope at Jodrell Bank in the UK, the Nan\c{c}ay Radio
Telescope in France, the Effelsberg Radio Telescope in Germany,
and the Westerbork Synthesis Radio Telescope in the Netherlands
(Table\,\ref{tab:tels}).
We have used the EPTA Data Release 1.0 (DR1; \citealp{dcl+16})
covering the time span MJD\,50931-56795 and combined this with TOAs
recorded using the Lovell telescope's analogue filterbank (AFB)
backend for the epoch MJD\,49030-55333, as well as some pre-DR1
Effelsberg TOAs using the Effelsberg-Berkeley pulsar processor (EBPP)
backend and the Effelsberg Pulsar Observing System (EPOS). The TOAs and ephemeris for this pulsar will be made available on the EPTA webpage \footnote{\url{http://www.epta.eu.org/aom/}}.

The AFB and EPOS data were aligned with the DR1 data set using the default procedure of fitting constant phase offsets between the data sets at each observing frequency, as described in \cite{dcl+16}. For a small subset of the AFB data at 1400\,MHz, known hardware configuration changes were corrected for by adding a phase offset to the corresponding TOAs.

\subsection{Jodrell Bank analogue filterbank data}
The AFB backend was used for
pulsar observations with the Lovell telescope during the years
1982-2010. TOAs were derived from observations at centre
frequencies of 400\,MHz, 600\,MHz, and 1400\,MHz, 
and a time resolution of $250\,\mu$s (see \citealp{hlk+04}). 
Observations were hardware-dedispersed and average profiles 
were produced via pulse folding. Each TOA was generated
through cross-correlation with an observing-frequency-specific
template, and systematic offsets between different instruments and observing configurations were corrected for by fitting for constant offsets.
The AFB data used a separate clock file for timing analysis (effectively treating the AFB as using a separate observatory to the digital filterbank used in more recent Lovell Telescope observations), as the AFB data had clock corrections already applied to the profiles, effectively absorbing the correction into the TOAs.

\subsection{Effelsberg Pulsar Observing System}
The EPOS backend (\citealp{j96}) recorded
observations using a 1390\,MHz centre frequency, with a 40\,MHz band
split into sixty 666\,kHz channels, which were digitally delayed
(incoherently dedispersed) to correct for the DM of the
pulsar. Observations were recorded at a time resolution of 60\,$\mu$s, and folded using early Jodrell Bank timing
models. The observations were timestamped using a local hydrogen maser
which was corrected to GPS. For more information on the EPOS system, see e.g.
\cite{kxl+98}.

\begin{table*}
\caption{PSR\,J0613$-$0200 timing data used in this study. The given RMS
    refers to the solution {\it including} the fit for the glitch as
    presented in Table\,\ref{tab:glitch} when the MJD range covers the
    glitch epoch. \label{tab:tels}}
\centering 
\begin{tabular} {c c c c c c}
\hline 
Telescope & Backend & Centre Freq. (MHz) & $N_{\text{TOAs}}$ & MJD Range & RMS ($\mu$s) \\
\hline
\hline
Effelsberg & EPOS & 1390 & 239 & 49768-51894 & 86.6 \\
\   & EBPP &1360 & 46 & 54483-56486 & 1.5 \\ 
 \  & EBPP & 1410 & 253 & 50362-54924 & 1.7 \\
 \ & EBPP & 2638 & 72 & 53952-56486 & 6.1 \\
\hline
Lovell & AFB & 400 & 132 & 49030-50696 & 47.7\\
\ & AFB & 600 & 142 & 49034-54632 & 22.3 \\
\ & AFB & 1400 & 586 & 49091-55333 & 18.3 \\
\ & DFB & 1400 & 24 & 54847-54987 & 5.4 \\
\ & DFB & 1520 & 191 & 55054-56760 & 2.0 \\
\hline
NRT & BON & 1400 & 334 & 53373-55850 & 1.1 \\
\ & BON & 1600 & 84 & 54836-56795 & 1.3 \\
 & BON & 2000 & 51 & 54063-56224 & 2.3 \\
\hline
WSRT & PUMA1 & 328 & 34 & 51770-55375 & 10.5 \\
\ & PUMA1 & 382 & 27 & 51770-55375 & 8.0 \\
\ & PUMA1 & 1380 & 99 & 51389-55375 & 3.0 \\
\hline
\hline
Total & - & 328-2638 & 2314 & 49030-56795 & 2.7 \\
\hline
\end{tabular}
\end{table*}

\begin{table*}
\caption{PSR\,J0613$-$0200 rotation parameters derived after fitting for the observed timing
  glitch, using the full EPTA Data Release 1.0 TOAs, pre-Data Release 1.0 EBPP 1410\,MHz TOAs, EPOS 1390\,MHz TOAs, and Jodrell Bank AFB
  400-1400\,MHz TOAs. Glitch parameters are estimated using our frequentist and Bayesian models described in the text.\label{tab:glitch}} \centering
\begin{tabular} {l c c c}
\hline 
Parameter & Frequentist Value & Bayesian Model (inc. sys. noise) & Bayesian Model (no sys. noise)\\
\hline
Frequency epoch (MJD) & 55000 & - & - \\
Frequency (Hz) &326.6005620227(2)& - & - \\ 
Frequency derivative (Hz/s) & $-1.0228(4) \times 10^{-15}$ & - & -\\
Glitch epoch (MJD) & $50888(30)$& 50874(25) & 50922(14)\\
Glitch frequency step (Hz) & $8.2(3) \times 10^{-10}$ & $8.7(6) \times 10^{-10}$ & $7.6(3) \times 10^{-10}$\\
Glitch frequency derivative step (Hz/s) &$ -1.6(39) \times 10^{-19}$ & $+1.1(65) \times 10^{-19}$ & $-1.2(4) \times 10^{-18}$\\
\hline
\end{tabular}
\end{table*}

\section{Results}
\begin{figure*}
	\includegraphics[scale=0.85]{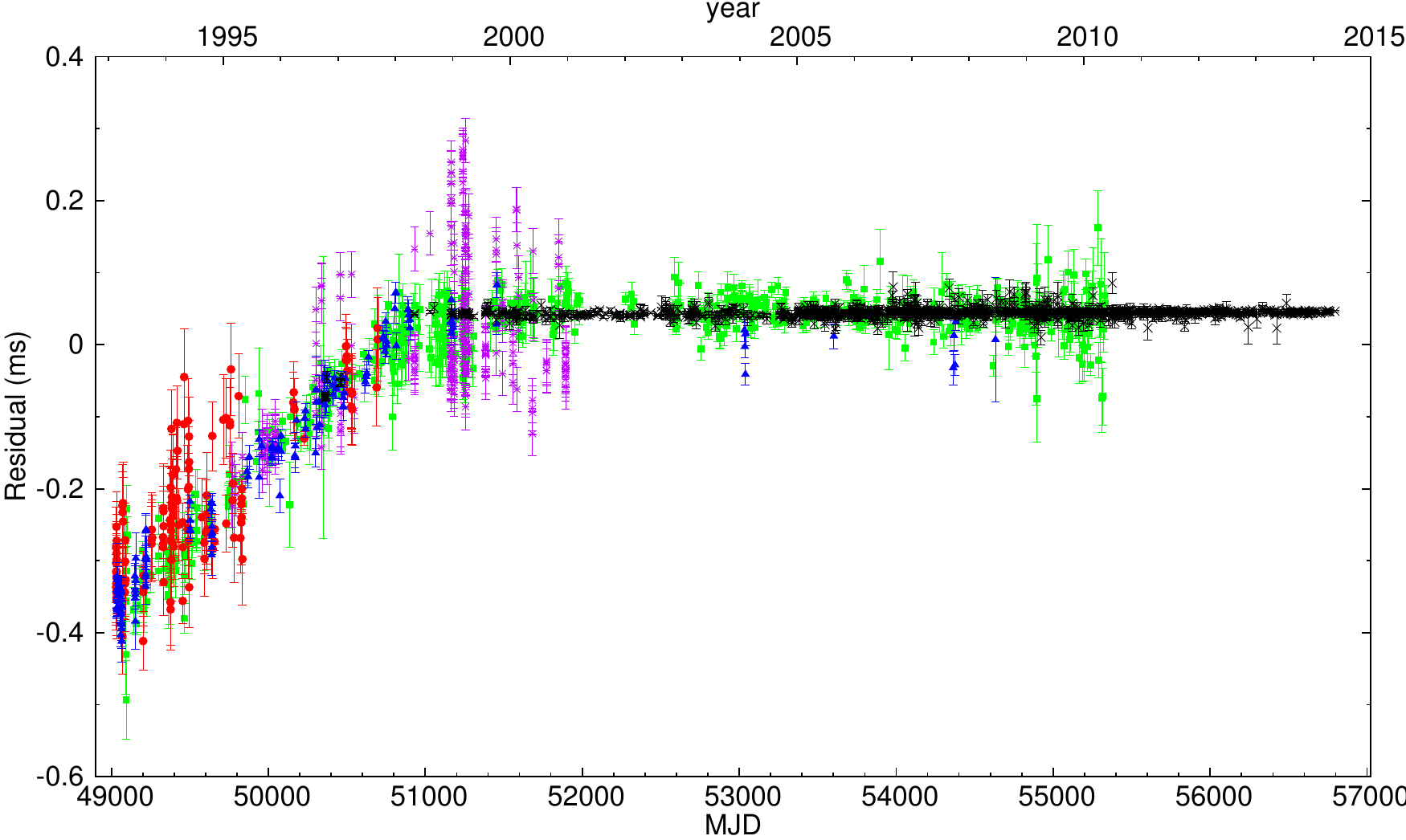}
	\centering
	\caption{Timing residuals for PSR\,J0613$-$0200 (\textit{purple stars}: EPOS 1390\,MHz \textit{green squares}: AFB 1400\,MHz \textit{blue triangles}: AFB 600\,MHz \textit{red circles}: AFB 400\,MHz \textit{black crosses}: DR1 including earlier EBPP 1410\,MHz). TOAs were recorded using the instruments listed in Table\,\ref{tab:tels}, and analysed based on the EPTA Data Release 1.0
          ephemeris. The timing solution does not accurately predict
          the arrival time of average pulses for dates earlier than
          the original EPTA data set (MJD\,50931-56795), due to an
          unmodelled timing glitch occurring shortly before the epoch
          over which the EPTA ephemeris was derived.}
	\label{fig:toas}
\end{figure*}
\begin{figure*}
	\includegraphics[scale=1.0]{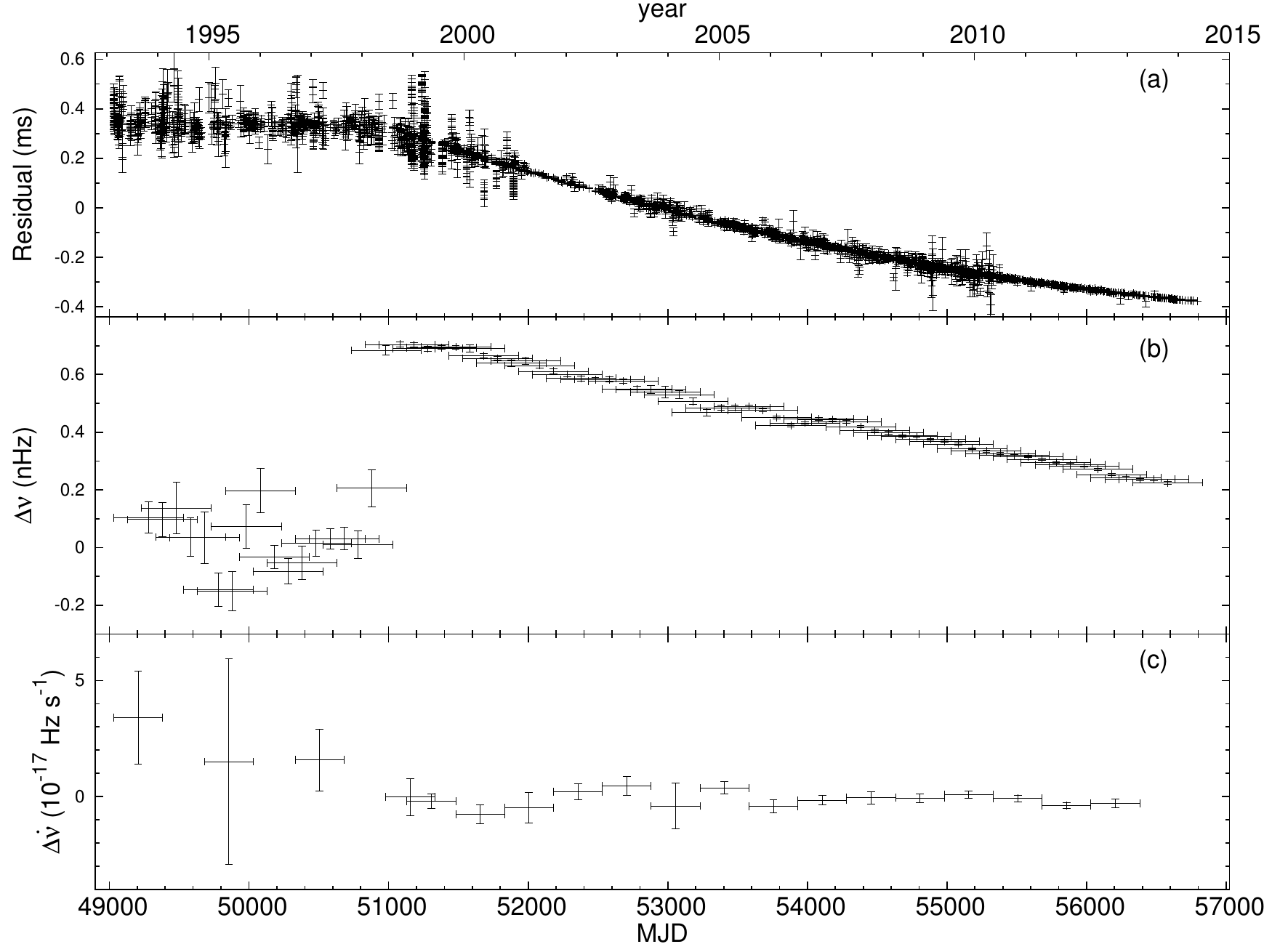}
	\centering
	\caption{Rotational frequency evolution for PSR\,J0613$-$0200
          from the pre-glitch ephemeris. (a) Timing residuals using a pre-glitch ephemeris, and our full data set described in Table\,\ref{tab:tels}. (b) Spin-frequency deviation from the ephemeris value derived using a stride fit through our full set of TOAs, using a 500-day window (represented by the horizontal error bars), and a 100-day stride length. (c) Frequency derivative variation from the ephemeris value over 350-day windows.
The discontinuity is caused by the timing glitch occurring at MJD\,50888(30) of fractional size $\Delta \nu/\nu=2.5(1) \times 10^{-12}$, and the linear decrease by a change in frequency derivative of fractional size $\Delta \dot{\nu}/\dot{\nu}=-1.6(39) \times
10^{-4}$ (both values measured by fitting for pre-glitch and post-glitch spin parameters using the entire data set).}
	\label{fig:f0}
\end{figure*}

Combining the DR1 and earlier AFB
TOAs revealed a sharp drift away from the DR1 timing solution, which was derived over the epoch
MJD\,50931 to 56795 (Figure \ref{fig:toas}). 
Comparing the TOAs of pulsars timed to similar or better precision over the same time span did not show any similar drift. This rules out the possibility of an instrumental effect or an error in the clock corrections as the cause of the drift seen in the early data for PSR\,J0613$-$0200.

TOAs recorded using the Effelsberg EBPP backend in the epoch MJD\,50362-50460 (i.e. preceding the start of DR1) and those from the EPOS backend were found to follow the same trend away from the predicted arrival
time as the Lovell Telescope AFB data,
excluding instrumental effects as the cause. ISM
effects such as a steadily changing dispersion measure can also be ruled out,
as the effect is present and identical in data from three widely-separated observing
frequencies without showing any frequency-dependent trend which would
be expected if the cause was ISM related.

The observed quasi-linear trend in the residuals is strong
evidence of a timing glitch, 
and can be removed completely by fitting for glitch parameters not
previously included in the timing solution, using a glitch epoch MJD\,50888 ($16^{\text{th}}$ March 1998), which allows the pre-glitch and
post-glitch frequency and frequency derivative to be derived (Table\, \ref{tab:glitch}). Fitting for the spin parameters before and after the measured glitch epoch, we measure the fractional frequency step to be ${\Delta
\nu/\nu=2.5(1) \times 10^{-12}}$, and the fractional frequency
derivative step to be ${\Delta \dot{\nu}/\dot{\nu}=1.6(39) \times
10^{-4}}$, where here and elsewhere we use a $1\sigma$ uncertainty. The change in spin frequency over time was investigated by
using a stride fit through our full data set, using a 500-day fitting
window, and a step size of 100 days, which was necessary for deriving precise values for the spin frequency from the relatively large uncertainties in AFB and EPOS TOAs, while still 
 allowing the
sudden change in $\nu$ to be clearly identified (Figure \ref{fig:f0}). 
The glitch epoch was estimated by fitting for all model parameters using \textsc{Tempo2} while varying the glitch epoch, and selecting the epoch corresponding to the minimum $\chi^{2}$. The uncertainty in glitch epoch was taken as the region over which varying the epoch results in ${\Delta \chi^{2}=1}$.
This is the smallest glitch ever recorded, with the next smallest also
occurring in a MSP, with fractional frequency step
${\Delta \nu/\nu=8(1) \times 10^{-12}}$ (\citealp{cb04}, \citealp{els+11})
i.e. 
several times larger than the glitch we report. 

The possibility of a magnetospherically-induced change in pulse shape related to a change in frequency derivative \citep{lhk+10} was considered as an alternative
to a change in spin frequency, but no significant change of the pulse
profile associated with the glitch was observed. 
However, it should be noted that the relatively low
time resolution of the AFB and EPOS observations is insufficient for small pulse-shape
changes to be detected.  
This effect was also tested for by fitting
for separate frequency derivatives only (i.e. no change in spin frequency)
for the pre-glitch and post-glitch residuals, using a range of
epochs for the change in frequency derivative while keeping the rest of the parameters constant. 
The glitch signature was not effectively
removed by 
this approach, with significant structure introduced to
the timing residuals. 
To remove the glitch signature, a model that includes a step in spin frequency is required, therefore we rule out magnetospheric effects as an explanation for this event.

Following the EPTA timing and noise analysis in \cite{dcl+16} and \cite{cll+16},
we use a Bayesian approach to confirm the findings of our frequentist analysis.
We estimate the properties of the glitch simultaneously with different noise models using the Bayesian pulsar timing package \textsc{TempoNest} \citep{lah+14}. These noise models include parameters to modify the properties of the white noise, as well as time-correlated stochastic signals that describe DM variations, timing noise, and system-dependent noise. For this final term, we use the approach described in \cite{lsc+16}. For all noise models, we marginalise analytically over the full timing model while simultaneously searching for a glitch epoch, and changes in the spin frequency and frequency derivative at that epoch. We use priors that are uniform in the glitch parameters, where the glitch epoch is the full MJD range of the data set, and the glitch frequency and frequency derivative priors are uniform in amplitude. All Bayesian evidence comparisons thus do not assume \textit{a priori} that a glitch is present in the data set. 

\begin{figure*}
	\includegraphics[scale=0.31]{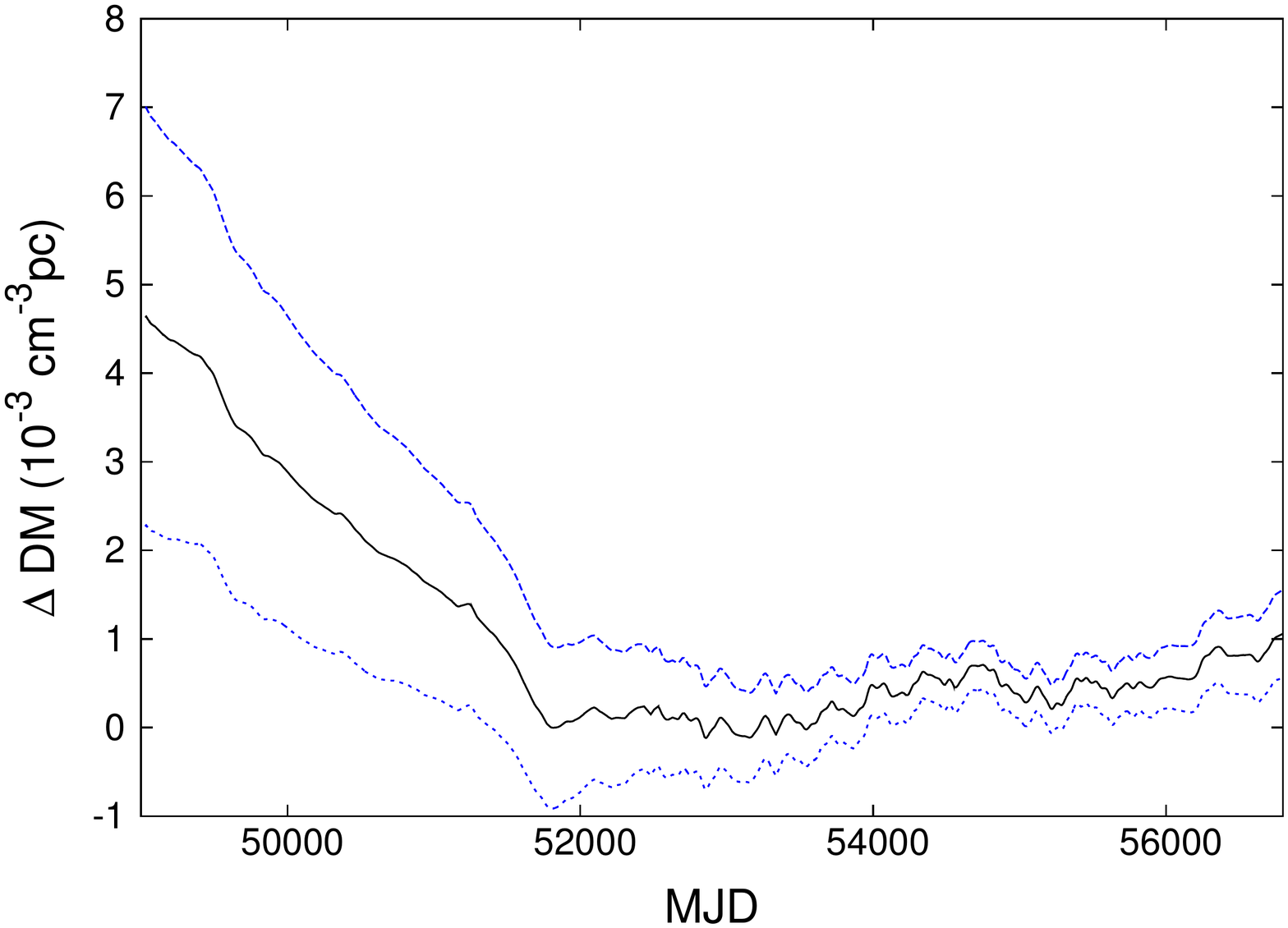}
	\includegraphics[scale=0.31]{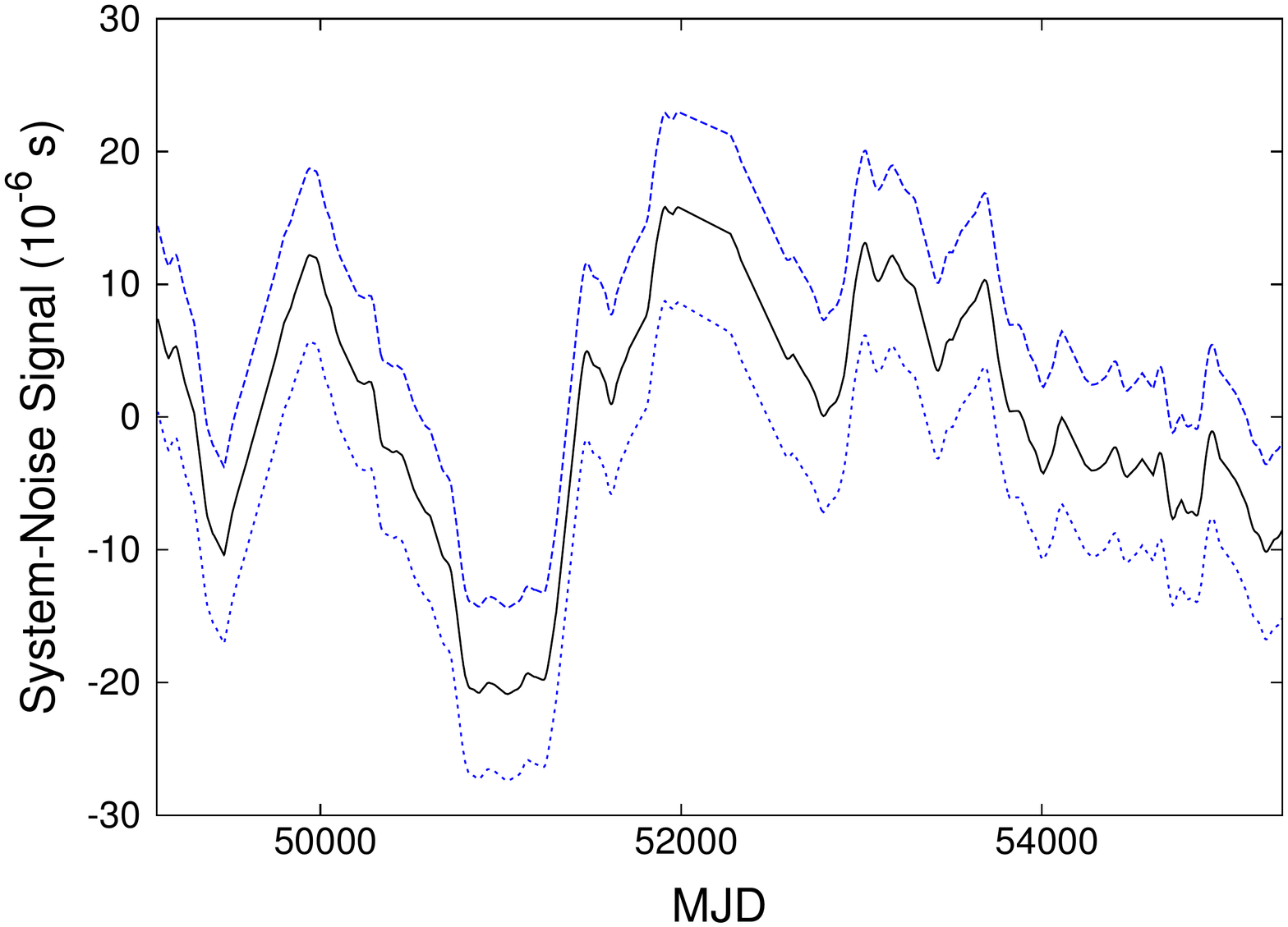}
	\centering
	\caption{The mean value (black line) and 1$\sigma$ confidence interval (blue lines) for DM variations (left panel), and system-dependent time-correlated noise in the AFB 1400\,MHz data (right panel) assuming a power-law model for both processes.}  
	\label{fig:GlitchPost}
\end{figure*}

We confirm the presence of a glitch, and find
a model that includes both DM variations and additional system noise in the AFB 1400\,MHz data set. 
We estimate a glitch epoch MJD\,50874(25) from the system noise model, and MJD\,50922(14) from the model without system noise. Using the system noise model, we estimate a spin-frequency step of 0.87(6)\,nHz and a spin-down rate step of 
${1.1(65) \times 10^{-19}\,\text{Hz} \ \text{s}^{-1}}$, and using the model with no system noise, we measure these quantities as 0.76(3)\,nHz and $-1.2(4) \times 10^{-18}\,\text{Hz} \ \text{s}^{-1}$ respectively.
In Figure~\ref{fig:GlitchPost}, we plot the mean signal realisation with 1$\sigma$ confidence intervals for the DM variations (top panel) and system noise (bottom panel) models. We find 
no evidence for a timing noise term that is coherent across all observing systems and is independent of the observing frequency (`spin noise' in \citealp{lsc+16}).
In principle, as we have only added additional observations to this data set compared to the DR1, we would expect that the sensitivity to 
timing noise
would either be the same or improve relative to that analysis. However, the presence of significant system noise in the early AFB 1400\,MHz data implies that the TOA estimates are affected by some time-correlated process that is potentially not well modelled by a stationary power-law noise process. If this early data were poorly modelled, then we would expect that including it in the data set would decrease our sensitivity to 
timing noise
compared to DR1 as observed. We test the stationarity of this system noise term by including two additional parameters that define the start time and duration of noise process. We find that the evidence does not increase with the addition of these parameters, with the start time consistent with the beginning of the AFB 1400\,MHz data set, and the duration consistent with the full length of the AFB 1400\,MHz data, implying that this system noise is not the result of mismodelling the glitch or a temporary increase in the noise level of the data set. However, there is not a sufficient overlap of data to distinguish the system noise term in this data set, as explained in \cite{lsc+16}.

\begin{figure*}
	\includegraphics[scale=0.6]{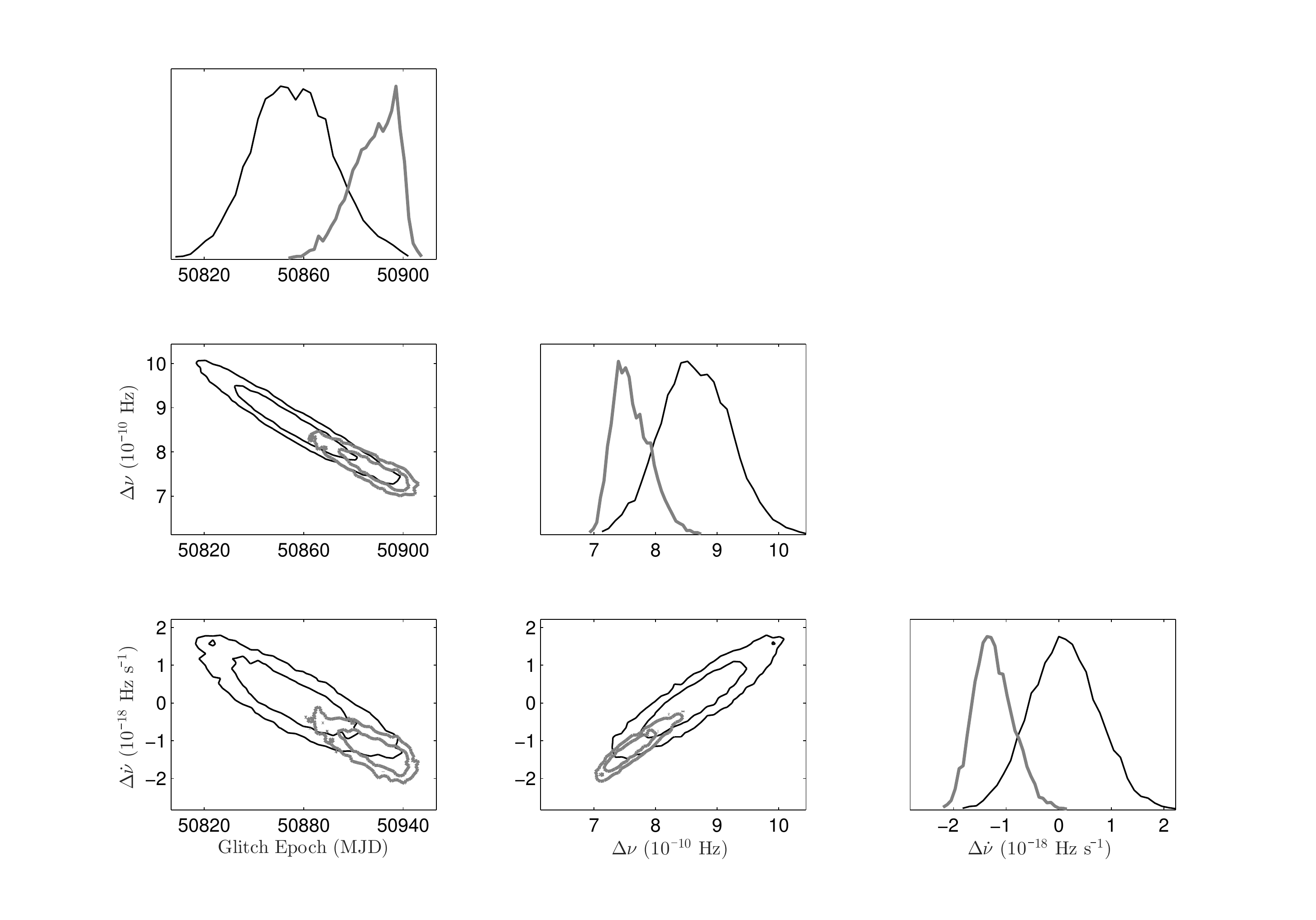}
	\centering
	\caption{One- and two-dimensional posterior probability distributions for the glitch epoch, the change in spin frequency, and the change in spin-down rate for two different models for the stochastic properties of the data set, with 1- and 2-$\sigma$ confidence intervals shown in the contour plots. \textit{Grey lines}: We include parameters that modify the properties of the white noise on a per system basis, scaling and adding in quadrature to the formal TOA uncertainties, and a power law model for the DM variations. \textit{Black lines}: We additionally include system-dependent time-correlated noise in the AFB 1400\,MHz data. We find the change in the log evidence is 24.7 in favour of the more complex model, indicating definitive support for their inclusion.}
	\label{fig:Signals}
\end{figure*}

In 
Figure~\ref{fig:Signals}, 
we show the one- and two-dimensional posterior probability distributions from our analysis for two different models. The black lines are from the 
optimal model
that includes system noise, DM variations, and white noise parameters. The grey lines are from an analysis that includes DM variations and white noise parameters only in the stochastic model. We find the increase in the log evidence for the model that includes system noise is 24.7, which definitively supports their inclusion in the model. 
We confirm the detection of the glitch and find that the the parameter estimates for the glitch model change significantly when including, or not, this additional system-dependent term. 
In particular the uncertainties in the change in frequency and spin-down rate increase by a factor of 1.8, and the mean of the change in spin-down rate is consistent with zero at the $\sim 0.2\sigma$ level compared to the greater than $3\sigma$ detection in the model without system noise, but the results are consistent with the results of the frequentist approach presented in Table \ref{tab:glitch}. 
 We stress that with these results we do not claim that the model for system noise used in the analysis is the most optimal that could be used. However, it is significantly preferred by the data compared to a model that does not include it at all.

\section{Discussion}
\subsection{Pulsar Timing Array Relevance}
The detection prospects for GWs using a PTA rely on the timing
of the pulsars included in the array to be extremely stable. Therefore
it may be reason for caution when a glitch is found in one of the most
stable MSPs included in current PTA projects.
However, our results show that the presence of a glitch in
PSR\,J0613$-$0200 does not affect timing stability for PTA analysis,
as TOAs used by the EPTA and IPTA for this pulsar are all derived
using post-glitch observations. The occurrence of a glitch before the
PTA epoch has not limited our ability to precisely time
this pulsar. This is shown by statistical analyses of pulsar
timing noise in PTAs, most recently by \cite{cll+16}, in which red noise is only semi-defined for this pulsar. As the glitch is small and the red noise of the pulsar is not well-defined, it is therefore
likely that
potential unmodelled glitches outside the timing baseline for other PTA pulsars have no significant effect on timing array sensitivity.

Including this work, only two glitches have been reported in MSPs, and one in the recycled pulsar B1913+16 \citep{wnt10}. 
Although small, the PSR\,J0613$-$0200 glitch was easy to detect with a data set covering a long
baseline. We can therefore be confident that no other glitches with similar sizes have been missed in the spin-evolution of pulsars observed at Jodrell Bank Observatory (JBO).
In this case, the effect of the glitch was easily removed without loss of timing precision, and so a glitch occurring in another PTA pulsar in the future may not be cause to remove the pulsar from future analysis. However, due to the unknown complexities of glitch models needed for MSPs, this is not completely certain. For future glitches in PTA pulsars,
only the pre-glitch data would be usable until sufficient time had passed for the post-glitch spin parameters to be measured, or for any post-glitch pulse profile
variation (\citealp{wje11}, \citealp{ksj13}) to be recognised in the case of a magnetospheric variation.


\subsection{MSP Glitch Rates}
Following the discovery of the first MSP glitch,
\cite{cb04} calculated an event rate of $\sim 1$ glitch per 500
pulsar-years of combined observations 
(or $\sim 0.2
\ \text{century}^{-1}$).
A total of 105 MSPs (period $P\,<\,10\,$\,ms) are observed at JBO, with a combined total of 1118 pulsar-years observing time. This allows us to estimate an event rate of $\sim 1$ glitch every 559 pulsar-years (or $\sim 0.18 \ \text{century}^{-1}$),
for glitches in MSPs of a size ${\Delta \nu/\nu \gtrsim 2 \times 10^{-12}}$, a rate consistent with Cognard \& Backer. 
We can extend this calculation to fully-recycled pulsars by including PTA pulsars with $P\,>\,10\,$\,ms and some double neutron star binaries (where we choose an upper limit of $P \sim 59$\,ms). We use this definition due to the difficulties in precisely defining recycled pulsars, but this allows us to estimate the order of magnitude of the glitch rate for this population. Following this, the glitch rate for recycled pulsars observed at JBO is $\sim 0.22 \ \text{century}^{-1}$.
We note that this rate is much lower than that for `normal' pulsars of 
$\sim 1$ glitch every 78 pulsar-years of observation (or $\sim 1.3 \ \text{century}^{-1}$). At JBO, 42 of the 49 IPTA pulsars (\citealp{vlh+16}) have been observed for a combined total of 793 pulsar-years, while the other 7 have 93.5 pulsar-years of observations in the IPTA data release. This gives a combined total of 886 pulsar-years, in which time only two glitches have been observed (a rate of $\sim 0.23 \ \text{century}^{-1}$). If we assume this is a good approximation of the PTA glitch return rate $r$ pulsar-years, then the probability of a glitch occurring in $t$ years is ${P=1-(1-r^{-1})^{t}}$. For observations of the 49 pulsars in the IPTA, this gives a probability of $\sim 70\%$ that another glitch will be observed in a PTA in the next 10 years, and $\sim 95\%$ that a glitch will be observed in the next 27 years. If we exclude PSR\,B1821$-$24A from the analysis, due to its unusual timing noise and acceleration within the host globular cluster, the return rate is $\sim 0.12$ century$^{-1}$, with a $\sim 50\%$ probability of a glitch in the next 10 years, and $\sim 95\%$ in the next $\sim 50$ years. As discussed earlier, the clear detection of such a small glitch in relatively low-precision data suggests that no such glitches have avoided detection in similarly precisely-timed MSPs.

It should 
be noted that the event rates calculated here assume that all MSPs and recycled pulsars are equally likely to experience a glitch, and that the probability remains the same following a glitch. This is probably not true, as the internal structure of the neutron stars is an important factor in the true glitch rate. The calculated glitch rates are therefore only estimates, but allow us to consider how likely it is that a glitch will occur in a $10+$ year data set required for a GW detection. 
There are also biases in our calculations which would need to be addressed for the true rates to be obtained. For example, we are biased against glitches occurring very early or late in a data set, due to the difficulty in recognising their effect.

\subsection{Neutron Star Structure}
The discovery of a glitch in PSR\,B1821$-$24A led to speculation on
the nature of neutron star structure, due to the small size of the
glitch, the high rotation frequency of the pulsar, and the
relatively low magnetic field strength (${2.25 \times 10^{9}}$\,G)
compared to other glitching pulsars ($\gtrsim 10^{11}$\,G).
\cite{mkd+09} interpreted this as evidence for PSR\,B1821$-$24A
being a strange star, due to the magnitude of the observed glitch
being consistent with the modelled values arising from a cracking of
the strange star crust. By comparison, we derive the inferred surface magnetic field strength of
PSR\,J0613$-$0200 from the period and spin-down rate to be ${1.7 \times 10^{8}}$\,G, an order of
magnitude lower than that of PSR\,B1821$-$24A. \cite{mkd+09} also
noted that the PSR\,B1821$-$24A glitch energy budget $\Delta E \sim 10^{40}$\,erg, given by ${\Delta E=\delta(I
\nu^{2}) \sim I \nu^{2}(\frac{\delta \nu}{\nu}) \sim
E_{\text{rot}}(\frac{\delta \nu}{\nu})}$, stood out from the rest of the population, which follow a line on a $\log \Delta E$ vs. $\log \Delta \nu/\nu$ plot, when assuming all
neutron stars have the same moment of inertia ${I=10^{45}\,\text{g}\,\text{cm}^{-2}}$. This implies that the large amount of energy required for such a change in angular momentum may not be readily available to millisecond pulsars. The
PSR\,J0613$-$0200 energy budget ${\Delta E \sim 2 \times 10^{39}}$\,erg
also does not follow the same distribution. It is therefore apparent that the combination of small
glitch sizes, greater characteristic ages (30\,Myr and 5\,Gyr for
PSR\,B1821$-$24A and PSR\,J0613$-$0200 respectively), lower magnetic
field strengths, and energy budgets imply that 
while MSPs are most likely neutron stars (e.g. recent MSP mass measurements in \citealp{ato+16}), they could potentially have
a different interior structure to the rest of the population,
which may cause the glitch mechanism or properties to be different. 
The uniqueness of the three glitching recycled pulsars can be seen in the $P$-$\dot{P}$ diagram (Figure \ref{fig:ppdot}).

\subsection{Gravitational Wave Memory}
One of the proposed causes of a GW signal in PTA data is a burst with memory (BWM),
caused by a merger of a supermassive black hole binary (SMBHB), which will leave a lasting change
(offset) in space-time \citep{bt87}. The main signature of such a burst in pulsar
TOAs is a step in frequency, without a step in frequency derivative. When a BWM passes over the Earth, a step will be seen at the
same time in all pulsars that are observed (Earth term). However,
since the signal travels at the speed of light, when a BWM passes
over a pulsar, it will not be seen in other pulsars in the PTA at the same
time, due to the large light travel time between pulsars. 
If a BWM affects only the pulsar term, this could be difficult to distinguish from a glitch, as only a single pulsar is affected.
There will also be no exponential recovery, as seen for some glitches \citep{cj12}. This would be difficult to identify in PSR\,J0613$-$0200, as \cite{lps95} noted the percentage glitch recovery decreases with characteristic age of the pulsar, making it effectively zero for a 5\,Gyr characteristic age.
\cite{mcc14} compare the BWM effect with the size of
the glitch in PSR\,B1821$-$24A. They conclude that if the frequency
change in that pulsar would have been caused by a BWM instead of a
glitch, this would have required an impossible scenario of a merger of
a $\sim 10^{10} M_{\odot}$ edge-on SMBHB only 10\,Mpc from the Milky Way. Such a system is excluded by single-source GW limits e.g. \cite{bps+16}, \cite{dec+15}, \cite{yss+14}.

We use the same argument to rule out a BWM scenario for the signature
in our data on PSR\,J0613$-$0200. Although the change in spin-down
rate is consistent with zero, the glitch size is too large to
make a BWM a realistic scenario for our measurements.

\section{Conclusions}
We have measured a spin frequency step in PSR\,J0613$-$0200 that we attribute to a small glitch, making this only the second detection of a glitch in a MSP, and the smallest glitch size recorded to date. We rule out other possibilities, such as magnetospherically-induced variations in rotation and pulse shape, and a gravitational wave BWM, due to the absence of effects associated with these causes. We interpret the difference between glitches in MSPs and the general pulsar population as potential indications of differences in MSP interior structure, and find that the glitch rate for MSPs is significantly different to that of the general population.
We demonstrate that glitch events are rare in PTA pulsars. Although their effect on the TOAs is significant, they can be accounted for without any further consequences for GW experiments when sufficient post-glitch timing data are available to correct for the glitch signature.

\section*{Acknowledgements}
We thank L.\,Levin for useful discussions. 
The authors acknowledge the support of the colleagues in the The European Pulsar Timing Array (EPTA). 

The EPTA is a collaboration between European institutes, namely ASTRON (NL), INAF/Osservatorio di Cagliari (IT), Max Planck Institut f{\"u}r Radioastronomie (GER), Nan\c{c}ay/Paris Observatory (FRA), University of Leiden (NL) and the University of Manchester (UK), with the aim to provide high precision pulsar timing to work towards the direct detection of low-frequency gravitational waves. An Advanced Grant of the European Research Council to implement the Large European Array for Pulsars (LEAP) also provides funding.

Part of this work is based on observations with the 100-m telescope of the Max-Planck-Institut f{\"u}r Radioastronomie (MPIfR) at Effelsberg.
Access to the Lovell Telescope is supported through an STFC consolidated grant.
The Nan\c{c}ay radio telescope is part of the Paris Observatory, associated with the Centre National de la Recherche Scientifique (CNRS), and partially supported by the R\'egion Centre in France.
The Westerbork Synthesis Radio Telescope is operated by the Netherlands Institute for Radio Astronomy (ASTRON) with support from The Netherlands Foundation for Scientific Research NWO.

S.O. is supported by the Alexander von Humboldt Foundation. P.L. gratefully acknowledges financial support by the European Research Council for the ERC Starting Grant BEACON under contract no. 279702.

This work was supported by the UK Science and Technology Research Council, under grant number ST/L000768/1




\bibliographystyle{mnras}
\bibliography{mjs+16v2} 


\bsp	
\label{lastpage}
\appendix
\section{\bf{$P$-$\dot{P}$} Diagram}
\begin{figure*}
	\includegraphics[scale=1.0,angle=270]{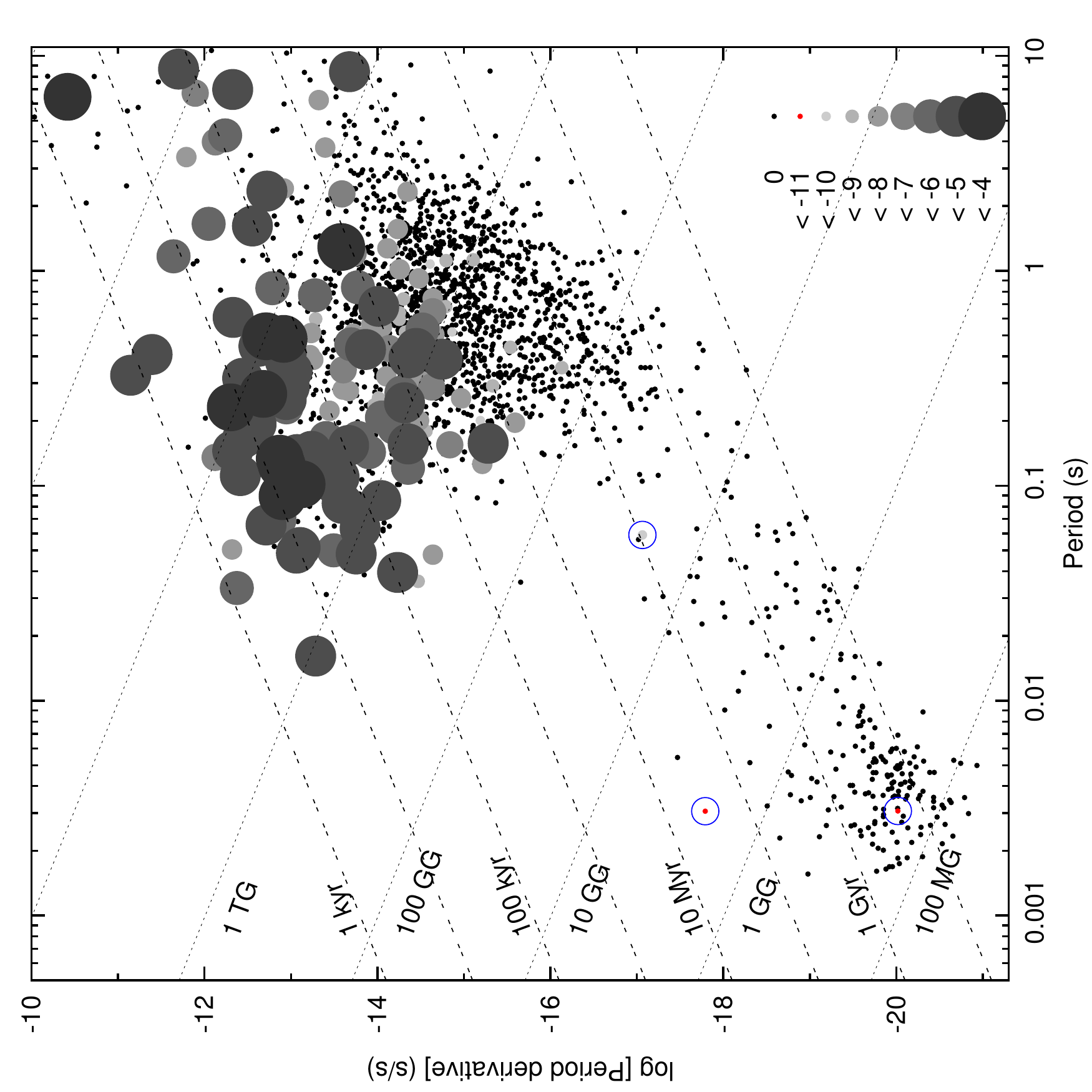}	
	\centering
	\caption{$P$-$\dot{P}$ diagram with pulsars labelled according to their cumulative spin frequency change due to glitches, $\log[\Sigma \,\Delta \nu/\nu]$. The labels in the key are the upper limit of the glitch size bins, which are in intervals of one order of magnitude. The three recycled pulsars for which a glitch has been observed are circled: B1913+16 (59\,ms, $8.6 \times 10^{-18}$\,s/s), B1821$-$24A (3\,ms, $1.6 \times 10^{-18}$\,s/s), and J0613$-$0200 (3\,ms, $9.6 \times 10^{-21}$\,s/s). The uniqueness of the PSR\,J0613$-$0200 glitch is apparent, due to its small size, and the location of J0613$-$0200 in the middle of the MSP `island'.}
	\label{fig:ppdot}
\end{figure*}
\end{document}